# Evolution of Microscopic Localization in Graphene in a Magnetic Field from Scattering Resonances to Quantum Dots

Suyong Jung[1,2*], Gregory M. Rutter[1*], Nikolai N. Klimov[1-3], David B. Newell[3], Irene Calizo[4], Angela R. Hight-Walker[4], Nikolai B. Zhitenev[1†], and Joseph A. Stroscio[1†]

[1]Center for Nanoscale Science and Technology, NIST, Gaithersburg, MD 20899
[2]Maryland NanoCenter, University of Maryland, College Park, MD 20742
[3]Electronics and Electrical Engineering Laboratory, NIST, Gaithersburg, MD 20899
[4]Physics Laboratory, NIST, Gaithersburg, MD 20899

**Graphene is a unique two–dimensional material with rich new physics and great promise for applications in electronic devices. Physical phenomena such as the half-integer quantum Hall effect and high carrier mobility are critically dependent on interactions with impurities/substrates and localization of Dirac fermions in realistic devices. We microscopically study these interactions using scanning tunneling spectroscopy (STS) of exfoliated graphene on a $SiO_2$ substrate in an applied magnetic field. The magnetic field strongly affects the electronic behavior of the graphene; the states condense into well-defined Landau levels with a dramatic change in the character of localization. In zero magnetic field, we detect weakly localized states created by the substrate induced disorder potential. In strong magnetic field, the two–dimensional electron gas breaks into a network of interacting quantum dots formed at the potential hills and valleys of the disorder potential. Our results demonstrate how graphene properties are perturbed by the disorder potential; a finding that is essential for both the physics and applications of graphene**.

The exposed and tunable two-dimensional graphene electronic system offers a convenient test bed for an understanding of microscopic transport processes and the physics of localization.

---

[*] These authors contributed equally to this work
[†] To whom correspondence should be addressed: nikolai.zhitenev@nist.gov, joseph.stroscio@nist.gov

Graphene's high transport carrier mobility and broad tunability of electronic properties promise multiple applications[1-3]. As in semiconductor devices, these features are ultimately determined by electron interactions and scattering from disorder including the surrounding environment of the device. Direct access to the graphene with scanned probes allows for the measurement of these interactions in greater detail[4-13] than possible in conventional semiconductor devices where the transport layers are buried below the surface. For example, STS with atomic resolution has been used[4,5] to study the local density of states of graphene and the role of disorder at zero magnetic field. Scanning single-electron transistor experiments, sensitive to local electric fields, produced local charge density maps with a spatial resolution of 150 nm[6] and detected single-electron charging phenomena at high magnetic fields[7].

In this article, we present STS measurements of a gated single-layer exfoliated graphene device in magnetic fields ranging from zero to the quantum Hall regime. With the ability to control the charge density of Dirac fermions with an electrostatic back gate with fine resolution, which was missing in previous STS studies[5,8-14], we can investigate local density of states and localization in graphene at the atomic scale while varying the Fermi energy ($E_F$) with respect to the Dirac (charge neutrality, $E_D$) point. At zero magnetic field, we observe density fluctuations arising from the disorder potential variations due to charged impurities underneath the graphene. At higher magnetic fields, discrete Landau levels (LLs) are resolved with both electron and hole states that follow single-layer graphene scaling. The LL spectra are dramatically different from previous STS measurements on epitaxial graphene on SiC[8,9,13] and graphene flakes on graphite[10,11], which are characteristic of weak disorder systems. Besides broader LLs due to disorder, we observe an additional set of localization resonances in the tunneling spectra, which are governed by single electron charging effects. Effectively, the localization in graphene can create a local quantum dot (QD)[7,15,16] with current flowing through two tunnel barriers in series; one barrier being the vacuum tunnel barrier between the probe tip and a local graphene QD, and the other barrier originating from resistive incompressible strips that isolate the QD in high magnetic fields. As a result, our STS measurements can not only detect local density of state variations in graphene with/without magnetic fields, but we are also capable of measuring the graphene electronic structure with sensitivity to probe single-electron charging phenomena at the Fermi level.

Figure 1a shows a scanning tunneling microscopy (STM) topograph of single-layer graphene exfoliated on $SiO_2$/Si substrate (see supplemental material) over an area of 60 nm × 60 nm. The peak-to-peak height corrugation of 1.2 nm is presumably due to the surface roughness of the underlying $SiO_2$[4,17,18]. As shown in the inset of Fig. 1a, the graphene honeycomb lattice is clearly resolved in atomic resolution STM images in any of the local areas.

First, we turn our attention to how charged impurities and structural disorder affect the local electronic properties of graphene in zero magnetic field. In an ideal graphene layer, the carrier density can be continuously tuned from hole to electron doping through zero density at $E_D$. However, as illustrated in Fig. 1c, local disorder gives rise to a spatially varying electrostatic potential that changes the relative position of $E_D$ with respect to $E_F$. By applying an external gate voltage, we can tune the level of the chemical potential, $E_F$ with respect to $E_D$, switching the charge carriers between electrons and holes. If $E_F$ is close to $E_D$, then spatially alternating patterns of electron and hole puddles are formed[4,6,19,20]. The variation of this local electrostatic potential has been considered as one of the main sources for the measured minimum conductivity in macroscopic graphene devices[20].

The distribution of the density set by the back gate voltage and spatially modulated by the disorder potential can be determined by locating $E_D$, a local minimum of the differential conductance $dI/dV$; a quantity that is proportional to the graphene local density of states. Figure 1d shows a representative $dI/dV$ spectrum with two distinct minima observed at tunneling bias voltages $V_b = -140$ mV and $V_b = 0$ mV. The minimum at $V_b = -140$ mV corresponds to $E_D$ while the zero-bias anomaly at $V_b = 0$ mV is characteristic of tunneling into graphene and other low-dimensional systems[5,21]. We note that the zero-bias anomaly in Fig. 1d appears different from that reported by Zhang et al.[14], where the spectral features were shifted off from $E_F$ by a large phonon gap. The zero-bias anomaly complicates an unambiguous $E_D$ determination when the average density is set close to the neutrality point. By setting the average $E_D$ away from zero bias and following the method developed in the reference 4, we obtain the spatial variation of $E_D$ from the mapping of $dI/dV$ at a fixed $V_b$ close to the average $E_D$ (Fig. 1b). The map provides an approximate landscape of the spatial distribution of the density fluctuations. Several distinct features with a characteristic length scale of 20 nm to 30 nm are identified by potential minima (pink colored region) and maxima (blue colored region). At low densities, these pink and blue areas would lead to electron and hole puddles, respectively.

While the approach described above provides a useful guide to identify the density fluctuations, accurate measurements of the fluctuations can be made only by finding the energy position of $E_D$ from the individual spectra at each spatial point. Tracking $E_D$ represented by a 'dip' in the $dI/dV$ map in the vicinity of $V_b \approx -140$ mV (Fig. 1e) yields a peak-to-peak variation of $\approx 20$ meV in a given disorder potential, which corresponds to a density fluctuation of $\approx 3 \times 10^{11}$ cm$^{-2}$. Near the second minimum in the $dI/dV$ spectrum in the vicinity of $E_F$ (zero bias), we observe a series of sharp resonances with spacing of the order of $\approx 15$ mV and linewidths of $\approx 10$ mV (Figs. 1d and 1e). We attribute these resonances to scattering from impurities or the disorder potential[5]. This interpretation is supported by the correlation between the spatial properties of the disorder potential and the spatial variation of the resonances (Figs. 1b and 1e).

More insight on the resonant peaks can be obtained from measuring *dI/dV* with respect to gate bias, i.e. gate mapping. The presence of the back gate controlling the charge density allows us to use a convenient approach to collecting and analyzing data in STS experiments as displayed in Figs. 2a and 2b. For a gate map, a series of *dI/dV* spectra is recorded as a function of both tunneling bias and gate voltages at a fixed spatial location. The Dirac point in zero magnetic field, represented by a *dI/dV* 'dip' in Fig. 2a varies in energy as a function of gate voltage. The dotted yellow line is from the best fit of the $E_D$ evolution with the Fermi velocity of $(1.12 \pm 0.01) \times 10^6$ ms$^{-1}$, which agrees well with those from previously reported exfoliated[6,14,22,23] and epitaxial graphene studies[8].

It is clear that resonant peaks, present around $E_D$, evolve in the same fashion as $E_D$ as a function of applied gate voltage (Fig. 2a), and persist even in low magnetic fields (Fig. 2b). We can easily identify that the resonance peaks are tied to the graphene dispersion relation, suggesting that the resonances are graphene derived states originating from impurity scattering or disorder potential, which have been a subject of recent theoretical interests in relation to the relativistic carriers of graphene[24,25]. These resonances would induce partial localization in the disorder potential, which will be reflected in magneto-transport studies.

In the presence of a perpendicular magnetic field, charge carriers in graphene circulate in cyclotron orbits with quantized energies referred to as Landau levels (LLs). As shown in Fig. 2b, the evolution of the graphene LLs can be tracked as a function of gate voltage; here at 2 T the evolution of the $N=0$ LL as a function of gate voltage is observed at the same location of $E_D$ following the identical dispersion relation as $E_D$ at 0 T (Fig. 2a). With increasing magnetic field, additional LLs develop along with the $N=0$ LL[8]. Figures 2c and 2d show a series of *dI/dV* spectra in magnetic fields up to 8 T at $V_g = -10$ V (hole carriers) and $V_g = 50$ V (electron carriers), respectively. As expected, the $N=0$ LL peak develops at the location of $E_D$ and additional LLs are distinctly resolved up to $N=7$ with both electron and hole carriers. It is instructive to point out that the LL linewidths are significantly broadened compared to previous reports on graphene on SiC[8,9], reflecting the shorter lifetimes and lower carrier mobility due to the substrate induced disorder potential. The LL energies follow the single-layer graphene scaling with energies that scale as $\sqrt{NB}$ and yield a carrier velocity of $(1.1 \pm 0.3) \times 10^6$ ms$^{-1}$ (see supplement material)[26]. This value is consistent with those obtained from the analysis of $E_D$-dependence on $V_g$ shown in Fig. 2

We now turn our attention to the regime where the $N=0$ LL is in the vicinity of $E_F$ in the quantum Hall regime. This low density regime corresponds to the breakup of the 2D electron system into electron and hole puddles due to the variations in the disorder potential[4,6,19,20]. Figures 3, a-c show gate maps of *dI/dV* spectra at 8 T at different spatial locations indicated by (a), (b), and (c) in Fig. 1b, respectively. The pinning of the LLs at $E_F$ produces a stair-case pattern of LLs in the gate maps[27-29]. Each LL is fixed at $E_F$ until all of its degenerate states are completely filled. Then, the next available LL is pulled to $E_F$, with other LLs followed correspondingly. Besides, a narrow dark band observed in the vicinity of $E_F$ is characteristic of a decrease in tunneling probability due to the development of a Coulomb gap in the quantum Hall regime[27].

An additional set of prominent features is visible in the gate maps appearing as bands of a quartet of peaks in the *dI/dV* spectra running diagonally. Note that the individual peaks only cross $E_F$ at which the particular LLs ($LL_0$, $LL_{\pm 1}$) are pinned. In addition, when the quartet of peaks intersect LLs at $E_F$, the respective LLs and the quartet of peaks form a well defined series of diamond structures (Figs. 3d and 3e), which are the main characteristic of Coulomb blockade physics. We identify the four-fold peaks as single-electron charging phenomena arising from Coulomb blockade effects in a graphene quantum dot (QD). A quartet of charging peaks has been observed with similar STS measurements in single-walled carbon nanotube QDs[30] and can be attributed to two valley-degenerate quantum states accommodating two electrons each (spin-up and spin-down).

Here we assert that graphene QDs in our device are defined by the underlying disorder potential and the incompressible (resistive) strips formed around potential hills and valleys in the quantum Hall regime (Fig. 4a). While the Coulomb charging of localized states is known to dominate the microscopic behavior at the Fermi energy[6,7,16], the appearance of the charging peaks overlapping with other LLs over the large energy range of the STS measurements (Figs. 3, a-c) is somewhat unexpected. At higher sample bias, there exist several conducting channels contributing to individual *dI/dV* spectra. As illustrated in the Fig. 4b, opening of a new conducting channel between the tip and sample Fermi levels appears as a peak in *dI/dV* spectrum. Specifically, the charging resonant peaks are only seen in tunneling spectra when the energy levels of QD are aligned to the sample Fermi level[15,30].

We note that the group of four *dI/dV* peaks seen at high magnetic fields emerges from a broad band present at lower magnetic fields ($B < 4$ T) (Figs. 3f and 3g) that even exists at 0 T (Fig. 2a). A set of *dI/dV* bands, as indicated with white arrows in Figs. 2a and 2b, is visible around $V_g = 20$ V where $E_D$ is close to $E_F$. The relation between this zero/low field band and the charging peaks at high fields is apparent; both bands correspond to resonance phenomena at the Fermi level. The bands at zero/low fields are likely caused by the scattering resonances discussed above while the quartet of charging peaks at high fields are from the resonance tunneling through quantized energy levels inside graphene QDs.

Detailed plots of $dI/dV$ spectra displaying the evolution of a quartet of charging peaks as the magnetic field increases are shown in Figs. 3f and 3g for $LL_0$ near $E_F$ (Fig. 3f), and for $LL_1$ at high sample bias (Fig. 3g), respectively. As dictated by Coulomb blockade physics[15], the sample bias spacing between individual $dI/dV$ peaks is the energy required to add an additional charge into the QD. We assume that the energy spacing is mainly determined by the charging energy ($E_C$) because the Zeeman energy ($E_Z \approx 0.92$ meV at $B = 8$ T) is much smaller than the observed energy spacing. The charging peaks start developing around 5 T and the peak spacing is observed to be independent of magnetic field; the variance in $E_C$ from 5 T to 8 T is around 1 meV (see supplement material).

We note that the spacing between charging peaks is uniform at the intersection with $LL_1$ but splits into two groups for $LL_0$. It is more obvious from the size variation of Coulomb diamonds revealing that the central diamond of the $LL_0$ is larger than other diamonds (Fig. 3d). In contrast, the size of Coulomb diamonds for the $LL_1$ shown in Fig. 3e is uniform. We suggest that the increased energy splitting between the second and third charging peaks is caused by the lifting of the valley degeneracy for the $N=0$ LL. The additional energy gap due to this symmetry breaking is estimated to be 10 meV at $E_F$ at 8 T. Thus, we can extract $E_C$ associated with $LL_0$ as $(16.4 \pm 0.6)$ meV[26], from the energy spacing for the first two and the last two charging peaks (see supplement material) in the vicinity of $E_F$. We can also calculate the size of the QD with $E_C$ and capacitances obtained from the gate maps (See Methods). Given $E_C = (16.4 \pm 0.6)$ meV and the gate map (Fig. 3b) taken at the spatial location of (b) in Fig. 1b, we can estimate the diameter of QD to be $(45 \pm 1)$ nm[26].

Detailed information on the LL spatial profile in the disorder potential and subsequent QD formation can be obtained by spatially mapping the local density of states ($dI/dV$ signal) in the quantum Hall regime. A three-dimensional $dI/dV$ dataset was taken with a sample bias varying from -300 mV to 300 mV and 0.4 nm spatial resolution at $B = 8$ T and $V_g = 20$ V. Similar to the characteristics of the gate map measurements, two distinct sets of features associated with either the density of states in LLs or the charging phenomena can be seen in the spatial maps. In Fig. 4c, the dominant pattern is defined by the spatial extent of the $LL_0$ at $E_F$ revealing the spatial location of the graphene QD in the location (b) of Fig. 1b; a compressible (conductive) $LL_0$ dot is encircled by an incompressible (resistive) region that defines the QD (Fig. 4a). The map in Fig. 4d, measured at high sample bias ($V_b = 300$ mV), shows four well-resolved concentric ring-like features representing the individual Coulomb charging peaks in the QD. As the tip moves to the center of the QD, the capacitance between the tip and the QD increases, sequentially inducing electron additions to the QD. It is clear from the maps that the spatial distribution of QDs formed in Fig. 4 is highly correlated with that of electron-rich puddles and incompressible strips with hole-rich ones (Fig. 1b). This correlation explains why the charging peak separation energies are relatively insensitive to changes in magnetic field (Figs. 3f and 3g); the QD sizes are defined by the disorder potential landscape. Different QDs are also observed in different regions of the disorder potential (see supplement material).

Charging of the QDs by the tunneling electrons is sensitive to the spatial properties of the local disorder potential (Figs. 3, a-c). The bands of charging peaks intersect with LLs at the edge of the flat pinned plateau near $E_F$. Interestingly, this behavior is different for disorder potential minima (Figs. 3a and 3b) and potential maxima (Fig. 3c). As the density is increased by changing $V_g$, the next LL starts to be populated in the minima while the surrounding area remains at an integer filling factor forming an incompressible strip. Hence, for potential minima, the QD is well defined at the beginning of the plateau. An opposite evolution occurs at potential maxima; as the density is increased, the surrounding regions reach integer filling factor before the center of the potential maximum, thereby spatially defining the boundary of the QD as the LL is filled and the single-electron charging is observed at the end of the plateau. We note that the LL-transitions between pinned regions in Fig. 3 have the same slopes as the Coulomb diamonds indicating that the transitions between LLs pinned at $E_F$ are defined by the same interactions and capacitances as the single-electron charging of the graphene QDs.

Our results demonstrate that the localization phenomena in graphene contribute to the STS spectra through two distinct set of features, one being through the density of states probed at the tunneling energy, and the other identified in the STS gate maps here, being conductance resonances at $E_F$. It has been well understood that the STS measurement can probe the scattering and the localization of wavefunctions leading to modulation of local density of states as a function of location and energy[5]. The sensitivity of the STS measurements to the low-energy physics at $E_F$ was believed to quickly decay when the tunneling energy exceeds the energies of interaction and localization. However, in the general case of systems with low density of states, tip potential effectively gates the sample and modifies the transmission at the Fermi level. As a result, Fermi-level physics such as interaction and localization can be probed at much higher energies, as demonstrated here. The identification of this new channel is afforded by the unique exposure of the graphene electronic system at the surface and the tunability of localization by the application of a magnetic field.

**Methods**

**Extracting Fermi velocity from gate maps.** The Dirac point in zero magnetic field varies in energy as a function of gate voltage as $E_D = \hbar c^* \sqrt{\pi n}$, where $c^*$ is the dispersion velocity of graphene, $n = \alpha |V_g - V_O|$ is the two-dimensional (2D) charge-carrier density induced by the applied gate potential $V_g$, $V_O$ is the shift of the Dirac point created by local intrinsic doping, and $\alpha$ is determined by the gate capacitance. With an insulating $SiO_2$ layer of 300 nm thickness, $\alpha$ is estimated from a simple capacitor model[22,23] to be $7.19 \times 10^{10}$ cm$^{-2}$V$^{-1}$. The dotted yellow line in Fig. 2a is the best fit of the $E_D$ evolution with $c^* = (1.12 \pm 0.01) \times 10^6$ ms$^{-1}$ and $V_O = (20.0 \pm 0.1)$ V[26].

**Calculating the size of graphene QD from gate maps.** It is straightforward to estimate the QD capacitance and the size from the Coulomb diamond features[15]. The single-electron transport through the graphene QDs is controlled by the voltages applied at the tip, contact to graphene, and the back gate through the capacitances of the vacuum gap ($C_d$), the incompressible strip ($C_s$), and the gate insulator ($C_g$), respectively (Fig. 4a). The ratio of capacitances obtained from the slopes of Coulomb diamonds in Fig. 3b around LL$_0$-pinned region, $V_g \approx 15$ V, is $C_d : C_s : C_g = 31 : 23 : 1$. The charging energy is determined by total capacitance ($C_{tot}$) of the QD, $E_C = e^2 / C_{tot}$, where $C_{tot} = C_d + C_s + C_g$. From the measured $E_C = (16.4 \pm 0.6)$ meV, we calculate the total capacitance of the QD, $C_{tot} = 9.8 \times 10^{-18}$ F and the gate capacitance, $C_g = 1.8 \times 10^{-19}$ F, yielding the QD diameter of $(45 \pm 1)$ nm[26].

**Acknowledgments** We would like to acknowledge M. Stiles and S. Adam for fruitful discussions and S. Blankenship, A. Band, and F. Hess for their technical assistance. We thank D. Davidovic and C. E. Malec for informing us of their unpublished work on tunneling in graphene.

**Author contributions** The graphene sample was fabricated by S.J. and N.N.K. and STM/STS measurements were performed by S.J., G.M.R., N.N.K. and J.A.S. The data analysis and preparation of the manuscript were performed by S.J., G.M.R., J.A.S., D.B.N and N.B.Z. The Raman spectroscopy measurements to confirm single-layer graphene flakes were performed by I.C. and A.R.H.

**Additional information** The authors declare no competing financial interests. Supplementary information accompanies this paper on www.nature.com/naturephysics. Reprints and permissions information is available online at http://npg.nature.com/reprintsandpermissions. Correspondence and requests for materials should be addressed to N.B.Z and J.A.S.

**Figure Captions**

**Figure 1: STM topography and STS *dI/dV* measurements at zero magnetic field. a,** STM topography image, 60 nm × 60 nm, of exfoliated single-layer grahene on SiO$_2$/Si substrate.

(inset) 3.5 nm × 3.5 nm atomic resolution image showing the graphene honeycomb lattice. Tunneling parameters: set-point current $I = 100$ pA and sample bias $V_b = -300$ mV for both images. **b,** Fixed-bias closed-loop *dI/dV* map ($V_b = -300$ mV, $V_g = 40$ V) over the same area as Fig. 1**a** revealing the spatial distribution of the disorder potential. **c,** Schematic diagram of the disorder potential variation as a function of spatial location. The relative position of the Fermi energy to the Dirac point can be tuned by an electrostatic potential from a back gate. **d,** *dI/dV* spectra taken at the top-most position of the white arrow in Fig. 1**b** showing a *dI/dV* minimum at the Dirac point. A second minimum occurs at the Fermi level (zero sample bias) along with a series of sharp resonance peaks indicated by the vertical tick marks. STS parameters: set-point current $I = 300$ pA, sample bias $V_b = -300$ mV, root-mean-square (RMS) modulation voltage 4 mV and gate voltage $V_g = 40$ V. **e,** A sequence of *dI/dV* spectra taken along the white line in Fig. 1**b** showing a number of resonances that vary with the disorder potential variation. The vertical axis is the distance along the line and the horizontal axis is the sample bias. The green solid line indicates the spatial Dirac point variation ($\Delta E_D \approx 20$ mV) along the white arrow in Fig. 1**b**. The *dI/dV* intensity is displayed in a color scale. STS parameters are the same as Fig. 1**d**.

**Figure 2**: **Gate and magnetic-field dependence of STS *dI/dV* spectra. a-b,** *dI/dV* gate maps taken at a fixed location (marked (b) in Fig. 1**b**) as a function of sample bias and gate voltage at 0 T and 2 T, respectively. The yellow dash-dotted lines show the evolution of the Dirac point at 0 T ($LL_0$ peak at 2 T) as a function of gate voltage. The scattering resonances observed in Figs. 1**d** and 1**e** are seen to follow the variation of the Dirac point. Broad *dI/dV* bands marked with white arrows in Figs. 2**a** and 2**b** are from the confinement by *p-n* junctions at lower magnetic fields and are evolving into a quartet of charging peaks at higher fields as seen in Fig. 3. **c-d,** Landau level spectra for various magnetic fields from 0 T to 8 T with hole ($V_g = -10$ V) and electron ($V_g = 50$ V) carriers, respectively. The *dI/dV* curves are offset for clarity. The various LL indices are indicated. The first three LLs ($LL_1$, $LL_{\pm 2}$, $LL_{\pm 3}$) are indicated by colored triangles in (Figs. 2 **b-d**). STS parameters: set-point current $I = 300$ pA, sample bias $V_b = -300$ mV and RMS modulation voltage 4 mV.

**Figure 3: *dI/dV* gate maps around the Dirac point in the quantum Hall regime**. **a-c,** High resolution *dI/dV* gate maps obtained at 8 T at the locations (a-c), respectively, indicated in Fig. 1**b**. Locations (a) and (b) correspond to disorder potential minima, and (c) to a maximum. **d-e,** Magnified images of the green-boxed regions in Fig. 3**a** showing Coulomb diamond features where charging lines intersect with $LL_0$ (**d**), and $LL_1$ (**e**), at the Fermi level. **f-g,** Four-fold Coulomb oscillation spectra measured for various magnetic fields from 0 T to 8 T at a fixed location (marked (b) in Fig. 1**b**) around the Fermi level for $LL_0$ (**f**) and around -220 mV for $LL_1$ (**g**) (see dotted yellow arrows in Fig. 3**b**), respectively. Note that $LL_0$ peak evolves at the Fermi level at lower magnetic fields (2 T to 4 T, in (**f**)) and Coulomb blockade effects overshadow the $LL_0$-peak development at higher fields (5 T to 8 T, in (**f**)). The unequal (**f**) / equal (**g**) spacing between charging peaks is clearly seen. The blue dots in Fig. 3**b** correspond to bias and gate voltages for the *dI/dV* maps in Figs. 4**b** and 4**c**. STS parameters: set-point current $I = 300$ pA, sample bias $V_b = -300$ mV and RMS modulation voltage 2 mV.

**Figure 4: Formation of graphene QDs in the quantum Hall regime. a,** Schematic of the breakup of the graphene 2DEG into interacting QDs (compressible regions) separated by insulating strips (incompressible regions) in the quantum Hall regime. Capacitances around the graphene QDs are defined in the main text. **b,** Schematic of single electron tunneling events through the graphene QD defined by two tunneling barriers (vacuum barrier and incompressible strip (I.S.) barrier) at high sample bias. Inside the QD, the Landau level pinned at the Fermi level is quantized into discrete charging levels. The incompressible strip barrier is not larger than the gap between LLs and peaks in *dI/dV* are observed when a new channel appears either at the STM tip Fermi energy (i.e. Landau level density of states) or at the Fermi level of sample (i.e. quantized states of graphene quantum dots). **c-d,** Spatial *dI/dV* maps, 60 nm × 60 nm, at $V_g = 20$ V and $B = 8$ T over the same area as in Fig. 1**a**. **c,** The map at the Fermi level ($V_b = 0$ mV) shows a compressible $LL_0$ region surrounded by incompressible strips, which gives rise to an isolated graphene QD. **d,** Map of charging peaks at higher sample bias, $V_b = 300$ mV. These maps correspond to the locations in the gate map indicated by the blue circles in Fig. 3**b**. STS parameters: set-point current $I = 300$ pA, sample bias $V_b = -300$ mV and RMS modulation voltage 4 mV.

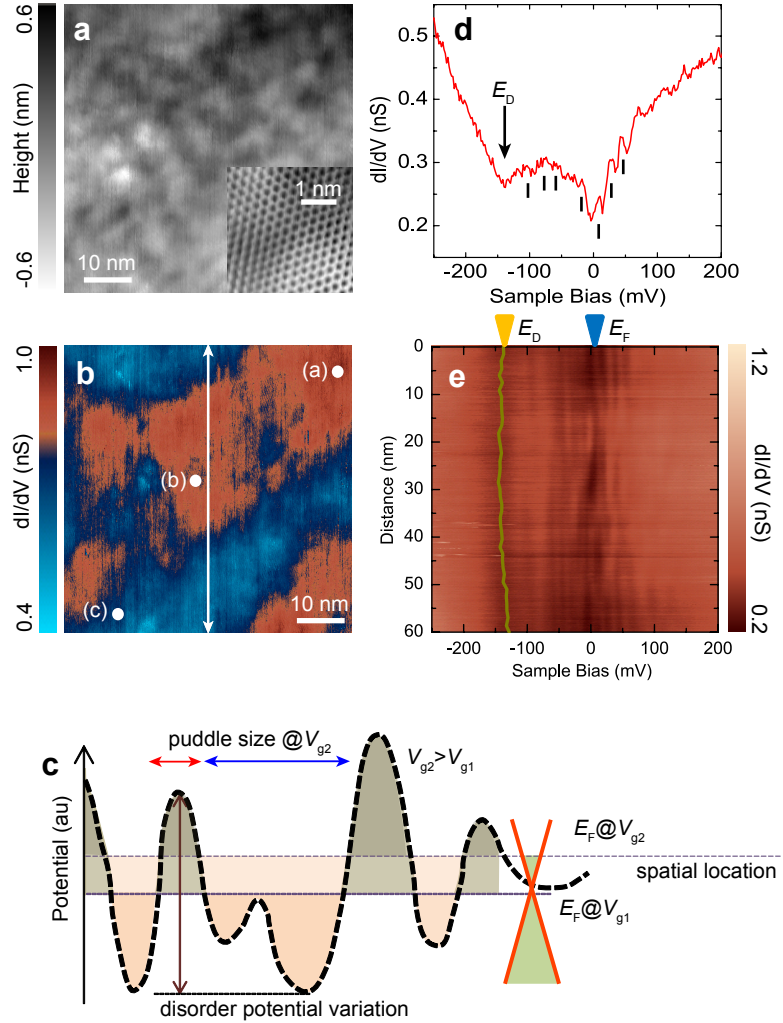

**Figure 1**: **STM topography and STS *dI/dV* measurements at zero magnetic field. a,** STM topography image, 60 nm × 60 nm, of exfoliated single-layer graphene on SiO$_2$/Si substrate. (inset) 3.5 nm × 3.5 nm atomic resolution image showing the graphene honeycomb lattice. Tunneling parameters: set-point current $I$ = 100 pA and sample bias $V_b$ = -300 mV for both images. **b,** Fixed-bias closed-loop *dI/dV* map ($V_b$ = -300 mV, $V_g$ = 40 V) over the same area as Fig. 1**a** revealing the spatial distribution of the disorder potential. **c,** Schematic diagram of the disorder potential variation as a function of spatial location. The relative position of Fermi energy to the Dirac point can be tuned by an electrostatic potential from a back gate. **d,** *dI/dV* spectra taken at the top-most position of the white arrow in Fig. 1**b** showing a *dI/dV* minimum at the Dirac point. A second minimum occurs at the Fermi-level (zero sample bias) along with a series of sharp resonance peaks indicated by the vertical tick marks. STS parameters: set-point current $I$ = 300 pA, sample bias $V_b$ = -300 mV, root-mean-square (RMS) modulation voltage 4 mV and gate voltage $V_g$ = 40 V. **e,** A sequence of *dI/dV* spectra taken along the white line in Fig. 1**b** showing a number of resonances that vary with the disorder potential variation. The vertical axis is the distance along the line and the horizontal axis is the sample bias. The green solid line indicates the spatial Dirac point variation ($\Delta E_D \approx 20$ mV) along the white arrow in Fig. 1**b**. The *dI/dV* intensity is displayed in a color scale. STS parameters are the same as Fig. 1**d**.

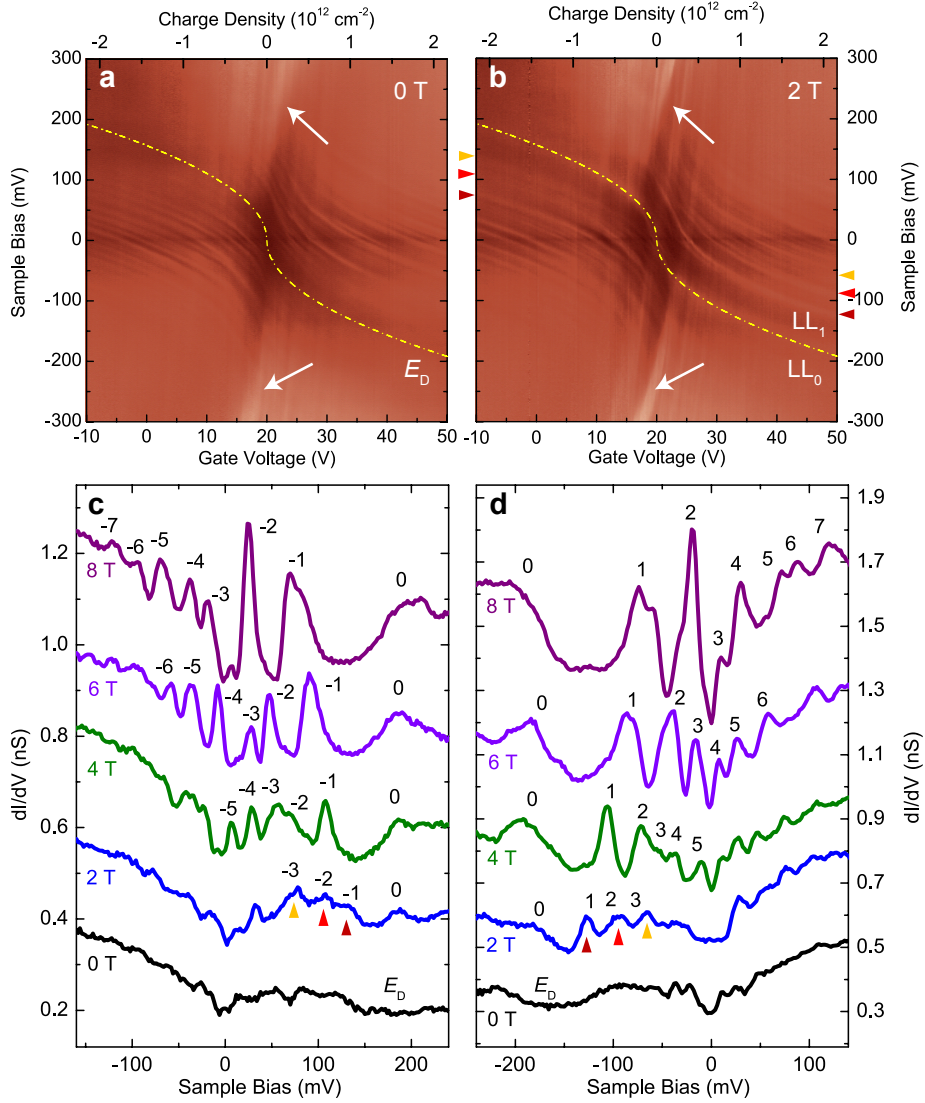

**Figure 2: Gate and magnetic-field dependence of STS *dI/dV* spectra. a-b,** *dI/dV* gate maps taken at a fixed location (marked (b) in Fig. 1**b**) as a function of sample bias and gate voltage at 0 T and 2 T, respectively. The yellow dash-dotted lines show the evolution of the Dirac point at 0 T (LL$_0$ peak at 2 T) as a function of gate voltage. The scattering resonances observed in Figs. 1**d** and 1**e** are seen to follow the variation of the Dirac point. Broad *dI/dV* bands marked with white arrows in Figs. 2**a** and 2**b** are from the confinement by *p-n* junctions at lower magnetic fields and are evolving into a quartet of charging peaks at higher fields as seen in Fig. 3. **c-d,** Landau level spectra for various magnetic fields from 0 T to 8 T with hole ($V_g$ = -10 V) and electron ($V_g$ = 50 V) carriers, respectively. The *dI/dV* curves are offset for clarity. The various LL indices are indicated. The first three LLs (LL$_1$, LL$_{\pm 2}$, LL$_{\pm 3}$) are indicated by color triangles in (Figs. 2 **b-d**). STS parameters: set-point current *I* = 300 pA, sample bias $V_b$ = -300 mV and RMS modulation voltage 4 mV.

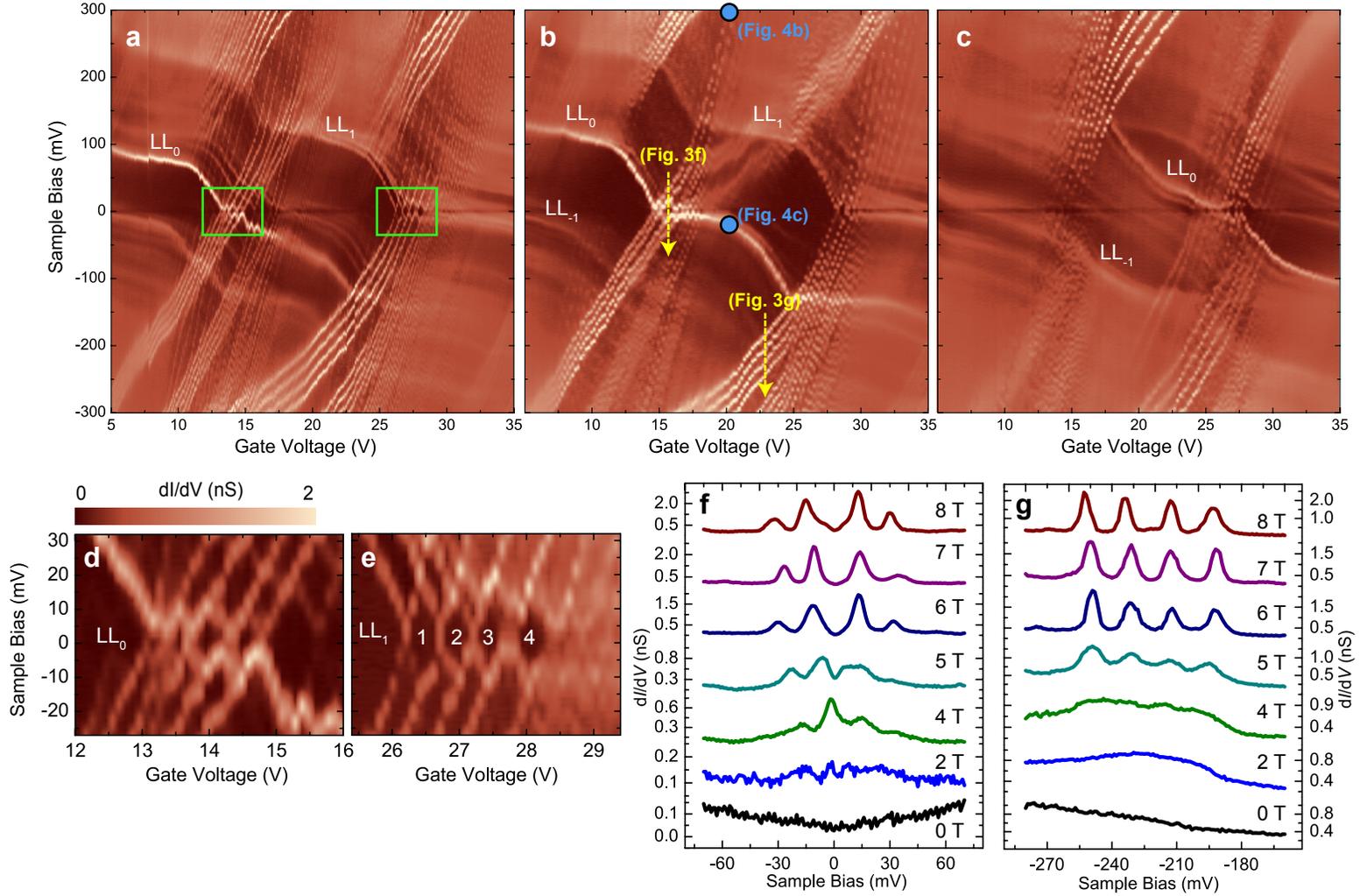

**Figure 3: *dI/dV* gate maps around the Dirac point in the quantum Hall regime**. **a-c,** High resolution *dI/dV* gate maps obtained at 8 T at the locations (a-c), respectively, indicated in Fig. 1**b**. Locations (a) and (b) correspond to disorder potential minima, and (c) to a maximum. **d-e,** Magnified images of the green-boxed regions in Fig. 3**a** showing Coulomb diamond features where charging lines intersect with $LL_0$ (**d**), and $LL_1$ (**e**), at the Fermi level. **f-g,** Four-fold Coulomb oscillation spectra measured for various magnetic fields from 0 T to 8 T at a fixed location (marked (b) in Fig. 1**b**) around the Fermi level for $LL_0$ (**f**), and around -220 mV for $LL_1$ (**g**) (see dotted yellow arrows in Fig. 3**b**), respectively. Note that $LL_0$ peak evolves at the Fermi level at lower magnetic fields (2 T to 4 T, in (**f**)) and Coulomb blockade effects overshadow the $LL_0$-peak development at higher fields (5 T to 8 T, in (**f**)). The unequal (**f**) / equal (**g**) spacing between charging peaks is clearly seen. The blue dots in Fig. 3**b** correspond to bias and gate voltages for the *dI/dV* maps in Figs. 4**b** and 4**c**. STS parameters: set-point current $I$ = 300 pA, sample bias $V_b$ = -300 mV and RMS modulation voltage 2 mV.

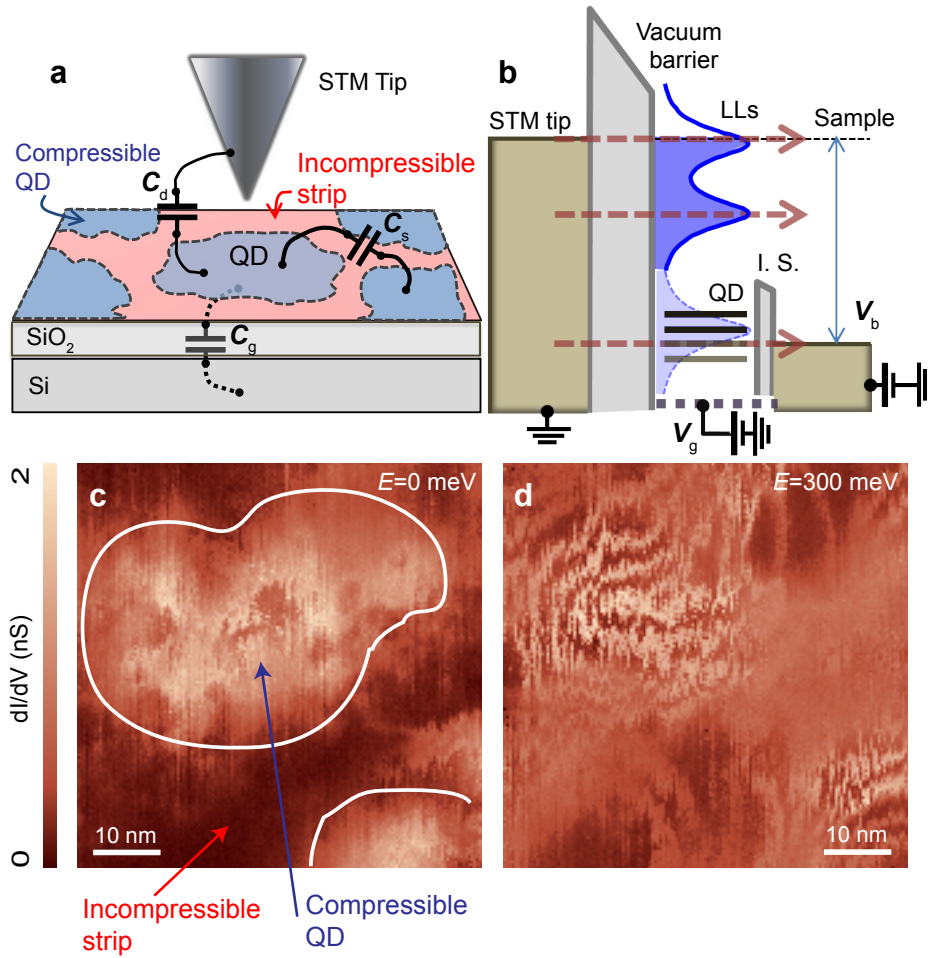

**Figure 4: Formation of graphene QDs in the quantum Hall regime. a,** Schematic of the breakup of the graphene 2DEG into interacting QDs (compressible regions) separated by insulating strips (incompressible regions) in the quantum Hall regime. Capacitances around the graphene QDs are defined in the main text. **b,** Schematic of single electron tunneling events through the graphene QD defined by two tunneling barriers ( vacuum barrier and incompressible strip (I.S.) barrier) at high sample bias. Inside the QD, the Landau level pinned at the Fermi level is quantized into discrete charging levels. The incompressible strip barrier is not larger than the gap between LLs and peaks in $dI/dV$ are observed when a new channel appears either at the STM tip Fermi energy (i.e. Landau level density of states) or at the Fermi level of sample (i.e. quantized states of graphene quantum dots). **c-d,** Spatial $dI/dV$ maps, 60 nm × 60 nm, at $V_g$ = 20 V and $B$ = 8 T over the same area as in Fig. 1**a**. **c,** The map at the Fermi level ($V_b$ = 0 mV) shows a compressible $LL_0$ region surrounded by incompressible strips, which gives rise to an isolated graphene QD. **d,** Map of charging peaks at higher sample bias, $V_b$ = 300 mV. These maps correspond to the locations in the gate map indicated by the blue circles in Fig. 3**b**. STS parameter: set-point current $I$ = 300 pA, sample bias $V_b$ = -300 mV and RMS modulation voltage 4 mV.

# Supporting Material for

# Evolution of Microscopic Localization in Graphene in a Magnetic Field: From Scattering Resonances to Quantum Dots

Suyong Jung, Gregory M. Rutter, Nikolai N. Klimov, David B. Newell, Irene Calizo, Angela R. Height-Walker, Nikolai B. Zhitenev*, and Joseph A. Stroscio*

\* To whom correspondence should be addressed:
Email: nikolai.zhitenev@nist.gov (N. B. Z.), joseph.stroscio@nist.gov (J.A.S.)

**This pdf file includes:**

- **I.** Materials and Methods
- **II.** Numerical Data Analysis of Gate Mapping
- **III.** Magnetic Field Dependence
- **IV.** Spatial Dependence
- **V.** STM-tip Related Features on STS Measurements

Figure S1-S6
Table S1 and S2
References

# Supplementary Material

## I. Experimental Methods

The experiments were performed with an ultra-high vacuum (UHV) scanning tunneling microscope (STM) facility at NIST in magnetic fields from 0 T to 8 T at a temperature of 4.3 K. The graphene device was fabricated in a similar way to that reported in Novoselov et al.[1]. Graphene flakes were mechanically exfoliated from natural graphite and transferred on thermally grown 300 nm thick $SiO_2$ on Si. The highly doped Si substrate was used as a back gate to control the charge density of the graphene device. Multiple steps of gold evaporation (50 nm for single deposition) through a SiN stencil mask were implemented to preserve a clean surface of graphene. Optical images of the graphene device are presented in Figs. S1a and S1b. A single-layer region located at the top-right corner is identified by lighter optical contrast and is confirmed by micro-Raman spectroscopy. A single-Lorentzian fit of the 2-D peak (Fig. S1c) verifies that the area of interest is indeed single-layer graphene[2].

The sample was thermally annealed at 140 °C in UHV for ≈ 18 h before being mounted into the microscope. An Ir probe tip was prepared by ex-situ electrochemical etching, and cleaned and characterized by in-situ field ion microscopy (FIM) before the measurements. (Detailed method for tip preparation will be discussed in the session V.) The tip was aligned over the sample at room temperature while being monitored through a long distance optical microscope. Finally, the STM was translated to the bottom of the cryostat cooling it down to 4.3 K. Figure S1a shows a schematic diagram of the sample, including an optical image of the graphene device with the STM tip after the alignment. A gate voltage is applied to Si substrate, and a bias voltage to the sample through a wire-bonded connection.

To locate the single-layer region under the STM, we first identified Au-contact step edges, and then positioned the tip inside the graphene area approximately 5 µm away from the Au edge to avoid additional doping or contamination from the Au electrode[3]. Scanning tunneling spectroscopy (STS) measurements were performed using a lock-in detection method with a modulation frequency of ≈ 500 Hz and root-mean-square (RMS) modulation voltages between 1 mV and 8 mV depending on the spectral range of interest.

## II. Numerical Data Analysis of Gate Mapping

The Dirac point at 0 T and the $N=0$ Landau level (LL) peak in weak magnetic fields follow the graphene dispersion relation:

$$E_D = E_{LL0} = \hbar c^* \sqrt{\pi n}, \qquad (S1)$$

where $c^*$ is the dispersion velocity of graphene, $n$ is the charge density, and $\hbar$ is Planck's constant divided by $2\pi$. The two-dimensional (2D) charge-carrier density, $n = \alpha|V_g - V_O|$ is defined by the applied gate potential $V_g$, and the shift of the Dirac point $V_O$, induced by intrinsic doping. With a 300 nm $SiO_2$ layer, $\alpha$ is estimated from the plain capacitor model[4,5] to be $7.19 \times 10^{10}$ cm$^{-2}$V$^{-1}$. Figure S2a shows the evolution of $LL_0$ peak measured at 2 T as a function of the gate voltage and its numerical fit to the Eq. (S1). The best fit shown by a solid red line yields $c^* = (1.12 \pm 0.01) \times 10^6$ m/s, and $V_O = (20.0 \pm 0.1)$ V[6]. The obtained velocity $c^*$ agrees well with those from the previously reported values for exfoliated[4,5,7,8] and epitaxial[9] graphene devices. The gate shift of 20 V corresponds to an intrinsic hole-doping of $1.44 \times 10^{12}$ cm$^{-2}$, which could have been introduced during the device fabrication.

The dispersion velocity can also be determined from the LL-spacing probed at high magnetic fields. The LL energies, $E_N$, of single-layer graphene follow the dependence on magnetic field ($B$) and LL index ($N$) as:

$$E_N = \text{sgn}(N) c^* \sqrt{2e\hbar |N| B}, \qquad (S2)$$
$$N = \ldots, -2, -1, 0, 1, 2, \cdots$$

where $e$ is the fundamental unit of charge. Figure S2b shows the experimental data and the fitting results at 6 T, yielding the values of $c^* = (1.12 \pm 0.1) \times 10^6$ m/s for hole carriers ($V_g = -10$ V) and $c^* = (1.07 \pm 0.3) \times 10^6$ m/s [6] for electron carriers ($V_g = 50$ V).

The analysis of the spectrum and the velocity determination at zero and low fields (Fig. S2a) has been performed assuming that the density is defined solely by the gate voltage. However, the potential difference between the tip and the sample certainly affects the local density at the tunneling point. This effect of the tip gating and the tip-to-sample capacitance is explicitly addressed in the analysis of the QD charging in the main text. In the general case, the tip-gate capacitance and the density profile is self-consistently defined by the 2DEG screening. In graphene, the screening is strongly density dependent, and the exact determination of tip-gating effect is not fully possible at the moment. However, the second determination of the Fermi velocity from the LL fan at high fields (Fig. S2b) is free error in the charge density. The close agreement between the velocity values determined by the two methods suggests that the error of the first method (Fig. S2a) is within the uncertainty of the second method (Fig. S2b).

### III. Magnetic Field Dependence

We have performed detailed magnetic field dependent measurements at the fixed location marked as (b) in Fig. 1b of the main text. Figures S3, a-d display a series of the gate maps obtained at different magnetic fields from 4.5 T to 7.5 T. The gate maps at 0 T (Fig. 2a), 2 T (Fig. 2b) and 8 T (Fig. 3b) are presented in the main text. It is evident that $LL_0$-peak evolution as a function of gate voltage significantly deviates from the Eq. S1 at $B = 4.5$ T and higher magnetic fields. At high fields, a clear stair-case pattern develops where LLs are pinned at $E_F$, followed by a sharp transition to another pinned level.

Another prominent feature seen in the gate maps at high magnetic fields is a quartet of oscillations running almost vertically. The oscillations sharpen as the $B$-field increases evolving into well-resolved peaks at high fields, as shown in Fig. S3e, which display individual $dI/dV$ spectra measured along the yellow arrows in Figs. S3, b-d. Using the Coulomb blockade model introduced in the main text, we can estimate how the QD size varies as a function of magnetic field. Table S1 shows the extracted charging energy and the estimated QD size at different magnetic fields from 5.0 T to 8.0 T. Here, the charging energy is the average value of the first two and the last two oscillation peaks in the selected range of gate voltage around the Fermi level. As shown in Table S1, the size of the observed graphene QD is insensitive to the applied magnetic fields, which suggests that the QD is not directly related to the magnetic length, which scales as $\propto 1/\sqrt{B}$, but is determined by the spatial properties of the disorder potential. Detailed analysis of charging energies for one of the QDs at 8 T is presented in Fig. S4.

### IV. Spatial Dependence

Figures S5 shows the spatial distribution of LLs obtained from three-dimensional maps of the local density of states as a function of charge density (gate voltage) and energy (sample bias). Within the 60 nm × 60 nm area of interest shown in Fig. S5a, we can identify four different QDs as illustrated in the closed-loop $dI/dV$ map in Fig. S5b. Figures S5, f-h are the subsets of the gate maps shown in Fig. 3 of the main text over a narrower range of gate voltage. Each gate map corresponds to individual QDs denoted by QD(a), QD(b) and QD(c) in Fig. S5b. Figures S5c and S5d show the spatial map of $N$=0 LL associated with the QD(a)/QD(d) and QD(b) at the gate voltage of 20 V and different sample bias voltages, respectively, with white color representing high density of states (large $dI/dV$ signal). The sample bias and gate voltage corresponding to the spatial $dI/dV$ maps is shown in the gate map with blue circles in Figs. S5f and S5g. The isolated pattern of $LL_0$ with the high density of states corresponding to each QD can be easily identified.

The quantum dots QD(a), QD(b) and QD(d) are formed at different minima of the disorder potential. In these three dots, $LL_0$ pinned at $E_F$ in the gate voltage range of 15 V to 20 V. In contrast, QD(d) is formed at a maximum of the disorder potential. As a result, $LL_0$ crosses $E_F$ in the vicinity of $V_g \approx 25$ V. Figure S5e shows the spatial location of the hole level, $LL_{-1}$, which is close to $E_F$ at $V_g = 10$ V (marked with the blue circle in Fig. S5h). Similar to cases above, the spatial correlation with the disorder potential maxima colored in blue (Fig. S5b) is clear.

The size of the QDs at the different locations is estimated from Coulomb blockade features; charging energy and the slopes of Coulomb diamonds, observed in the gate maps as described in the main text and listed in the table S2. Graphene QDs vary in size of 30 nm to 50 nm and have a strong correlation with the landscape of the disorder potential.

### V. STM-tip Related Features on STS Measurements

It is possible to observe anomalous tip-related states and tip effects in STS measurements, because STM/STS measurements are extremely sensitive to the surface states and the cleanness of the probe tip and samples. There have been several reports of tip-induced QDs in conventional 2DEG systems and the detailed mechanism of their QD formations and electronic properties has been well understood[10]. Recently, LeRoy *et al.*[11] reported that their STS data on graphene can be explained by tunneling through a QD attached to the tip.

In this section of the supplementary material, we detail: a) our probe tip preparation procedures, and b) how a QD can be formed on the probe tip by picking up a Au nanoparticle, and we show the dramatic differences in STS measurements with a QD on the probe tip vs. the QDs probed in the graphene 2DEG.

### a. Probe tip preparation

Our Ir probe tips are electrochemically polished under an optical microscope before *in-situ* cleaning with FIM. The tips typically have radii in the range of 5 nm to 10 nm and are imaged in a He image gas with tip potential of up to 10 keV. Prior to FIM imaging, the tip is heated by electron-beam heating to remove any chemical etching residue. The tip is then imaged in the FIM, and the applied potential is increased to field-evaporate the outer tip layers, until a high symmetry pattern characteristic of Ir is obtained. If a pattern does not develop before 10 keV (indicating a dull tip), the tip is discarded, and a new one is selected. This guarantees the probe tip is clean and has a radius of 5 nm to 10 nm. Note that these tip radii are much smaller than the size of the graphene QDs we measure, as detailed in the manuscript.

Our probe tips remain relatively clean in our measurements since our device fabrication avoids any wet-chemistry procedures, which guarantees no organic polymer or solvent residues exist on the graphene surface. It is also unlikely that tip contamination could be caused by detached graphene flakes from the graphene device. Our sample is made from a single continuous flake of graphite exfoliated on insulating $SiO_2$ surface (The optical image of fabricated device is shown in Fig. S1). No loosened flakes were encountered while scanning across the device area, and no bits of graphene layer were removed or torn apart during the scanning. The coverage by the single layer graphene was complete without any cracks in the graphene, as the tip would crash on the area over which the graphene film was removed and the $SiO_2$ surface exposed. During scanning, tip resolution changes as it is scanned repeatedly over the surface, which is a signature of rearrangement of the atoms at the end of the probe tip. To improve the resolution and stability, we positioned the tip over the Au electrodes and tunneled under field emission conditions (higher voltages) until the microstructure of the tip was stabilized. We have obtained thousands of spectra with different microstructures on the tip. The reproducibility of the data and the appearance of the QDs in the graphene 2DEG therefore cannot be due to an anomalous tip condition.

### b. Picking up Au nanoparticles

One byproduct of scanning over graphene close to the Au electrodes is the possible pickup of a Au nanoparticle that has been loosened from nearby electrodes. When this occurs, a QD can be formed on the tip apex, with dramatic changes in STS measurements. These STS spectra are completely different from spectra without a QD on the tip. In addition, when the QD formed on the tip it can be easily modified or removed by applying higher potentials to the probe tip or simply scanning over clean graphene or Au surfaces. In contrast, the spectra reported in the manuscript are obtained with a tip that does not change over time.

Supplementary Fig. S6 shows an example of the STS spectra with a QD attached to the tip at different gate voltages, which are drastically different from the spectra obtained on a QD formed in graphene. The STM tip picked up the QD during the searching for the single layer graphene region. Although the detailed origin of the QD formation remains unknown, it is likely due to the pickup of a Au particle, which are seen in the near the Au electrodes. The spectra in Fig. S6 are dominated by the regularly spaced Coulomb charging peaks of a QD. There are several notable differences in STS spectra from the graphene QD induced in the disordered potential in quantum Hall regime: (1) The values of charging energies are quite different. The energy spacings between the adjacent Coulomb peaks (as large as 100 meV) are much larger (a factor of 5 to 10 times) than those from the graphene QDs, reflecting the smaller size of the QD attached to the STM tip. (2) Coulomb oscillations are ubiquitous as a function of bias and gate voltages. No correlation with the electronic structures of underlying graphene devices such as the position of Dirac point and the Fermi level or the formation of Landau levels is observed. (3) The gate voltage dependence of the charging peaks of the QD on the tip is quite different compared to those from graphene placed on the $SiO_2$ surface because of the screening by the graphene and the additional vacuum gap between the QD on the tip and the back gate. (4) The charging peaks of the QD attached to the tip do not depend on magnetic field. (5) The microscopic configuration is unstable and the consistency of measurements is lacking. In general, QDs attached to the tip are loosely bound so that the STS spectra are easily influenced by the scanning conditions. We saw the Coulomb oscillations, as displayed in Fig. S6, change their positions in sample bias and/or completely disappearing even after short STM scanning over the surface.

To ensure the experimental data are free from the spurious tip effects as discussed above, we have paid special attention to the tip treatment, as described above. If Coulomb blockade from QDs on the tip were encountered in the STS spectra, we deliberately moved the tip on the Au region and treated the tip with field emission cleaning (a common method in STM) until the QD features were completely removed and an image of the clean Au surface was recovered. After reconditioning the microstructure of the tip, we started new measurements at a fresh graphene sample location. Therefore, the reproducibility and fidelity of the measurements in the manuscript and supplemental material is clear, and cannot be due to a QD on the probe tip.

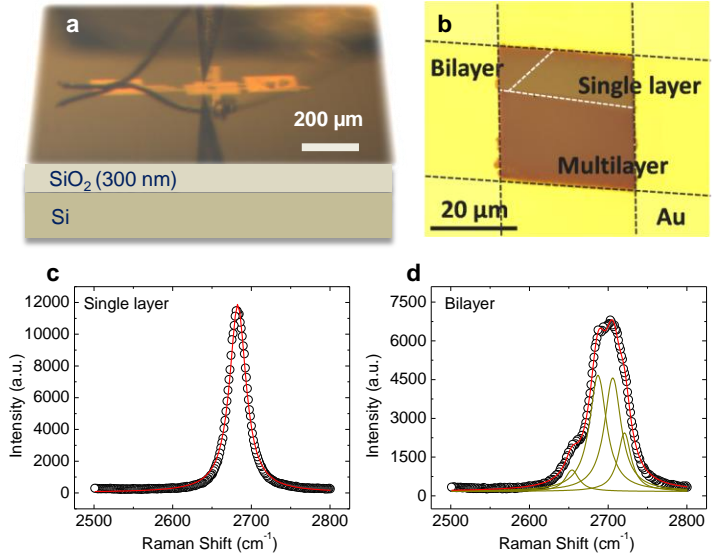

**Figure S1**: **a,** Optical image of the final graphene device with a STM tip aligned and a schematic of gate insulator ($SiO_2$) and a gate electrode (Si). **b,** Optical image of the device showing the graphene area of interest and Au step edges from contact evaporations. **c-d,** Micro-Raman spectroscopy data (open circles) and single Lorentzian fit (solid red line) of the single-layer graphene (**c**) and four Lorentzian fit (solid dark yellow lines) of the bilayer graphene (**d**).

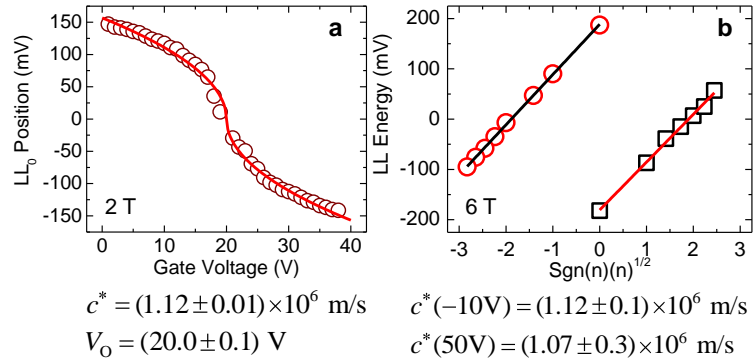

$c^* = (1.12 \pm 0.01) \times 10^6$ m/s
$V_O = (20.0 \pm 0.1)$ V

$c^*(-10V) = (1.12 \pm 0.1) \times 10^6$ m/s
$c^*(50V) = (1.07 \pm 0.3) \times 10^6$ m/s

**Figure S2: a,** Numerical fitting of $LL_0$-$V_g$ dependence at $B = 2$ T. The position of $LL_0$ (open circles) is extracted from the gate map presented in Fig. 2**b** of the main text. The least square fit using Eq. S1 is shown by red curve. **b,** Determination of graphene dispersion velocity from LL-peak positions at 6 T. Open circles and squares represent data extracted from individual STS spectra presented in Fig. 2**c** for holes ($V_g$ = -10 V) and Fig. 2**d** for electrons ($V_g$ = 50 V), respectively, and solid red lines are the least square fits.

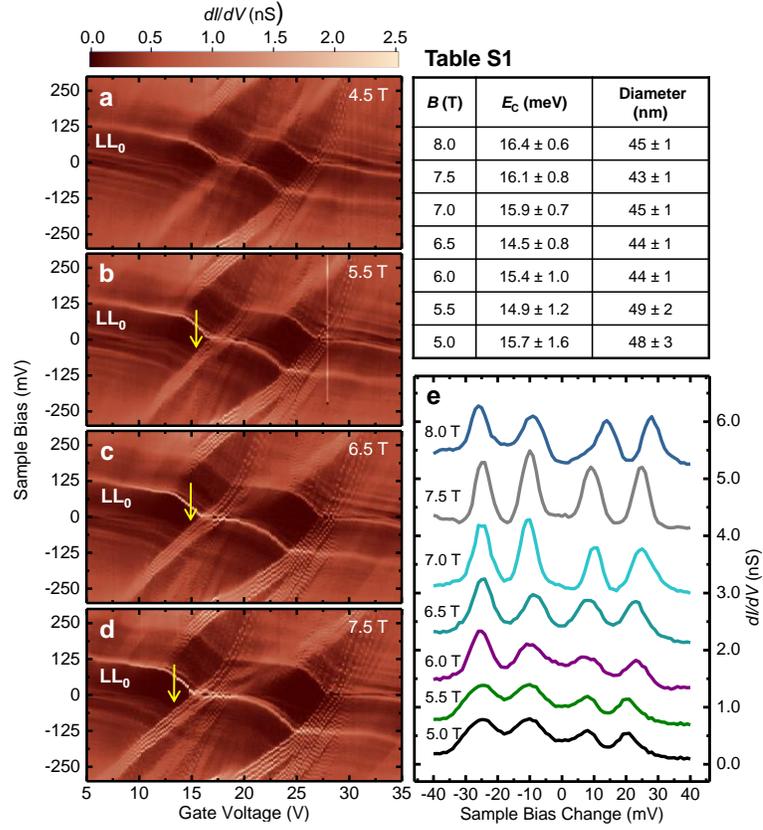

**Figure S3**: **a-d**, *dI/dV* gate maps obtained at the location (b) indicated in Fig. 1**b** in the main text at different magnetic fields. **e,** Four-fold Coulomb oscillation spectra measured at various magnetic fields from 5 T to 8 T around the Fermi-level for $LL_0$ ($V_g \approx 15$ V as indicated with a yellow arrow in Figs. S3, **b-d**). *dI/dV* spectra are offset for clarity.

**Table S1:** Charging energies determined by the energy spacing between the first two and the last two charging peaks and the estimated size of the QD at different magnetic fields.

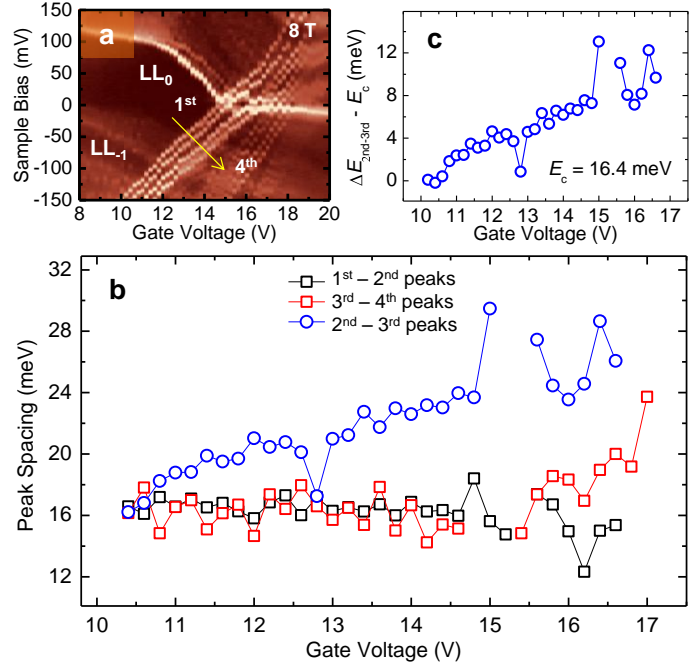

**Figure S4**: **a,** *dI/dV* gate map in the selected range of sample bias and gate voltages of Fig. 3**b** in the main text. The index of Coulomb charging peaks is labeled as indicated by the yellow arrow. **b,** Peak spacing at different gate voltages around the Fermi level. Each spacing is calculated from the obtained peak position by numerical fitting. Charging energy ($E_C$) from the spacing between the first two (1st and 2nd, black open squares) and the last two (3rd and 4th, red open squares) oscillation peaks remains constant in the gate voltage range of 10 V to 15 V. Calculated charging energy is (16.4 ± 0.6) meV. In contrast, the energy spacing between the 2nd and 3rd charging peaks (blue open circles), which is due to the lifting of the valley degeneracy, increases as charging peaks and $LL_0$ move closer to the Fermi level. **c,** The valley energy gap, determined from the energy spacing of 2nd and 3rd peaks after subtracted with charging energy, $E_C$ = 16.4 meV, as a function of gate voltage.

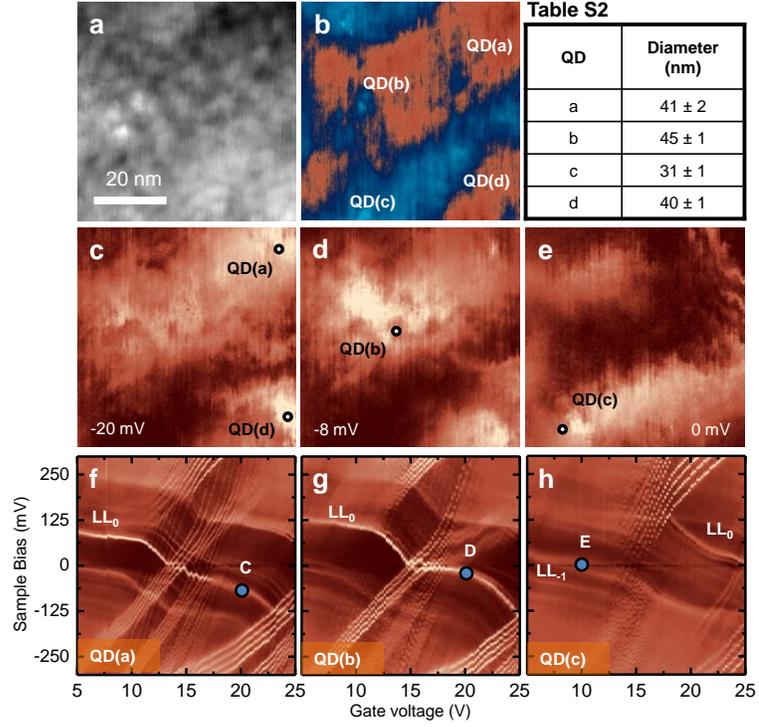

**Figure S5**: **a-b,** STM topography image and a fixed-bias closed loop $dI/dV$ map of 60 nm × 60 nm of single-layer graphene described in the main text. Locations of different QDs are marked in Fig. S5**b**. **c-e,** Spatial $dI/dV$ maps revealing compressible LLs that corresponds to individual QDs. Maps of $LL_0$ associated with QD(a) and QD(d) at $V_g$ = 20 V and $V_b$ = -20 mV (Fig. S5**c**), and QD(b) at $V_g$ = 20 V and $V_b$ = -8 mV (Fig. S5**d**). Map of $LL_{-1}$ associated with QD(c) taken at $V_g$ = 10 V and $V_b$ = 0 mV (Fig. S5**e**). **f-h,** $dI/dV$ gate maps measured at 8 T at the location (a), (b) and (c), respectively, in Fig. 1**b** of the main text. These maps are associated with the QD(a), QD(b) and QD(c) as indicated in Figs. S5, **c-e**. The blue dots indicate where spatial $dI/dV$ maps (Figs. S5, **c-e**) are taken in the map of sample bias and gate voltage.

**Table S2:** Estimated QD size at different locations.

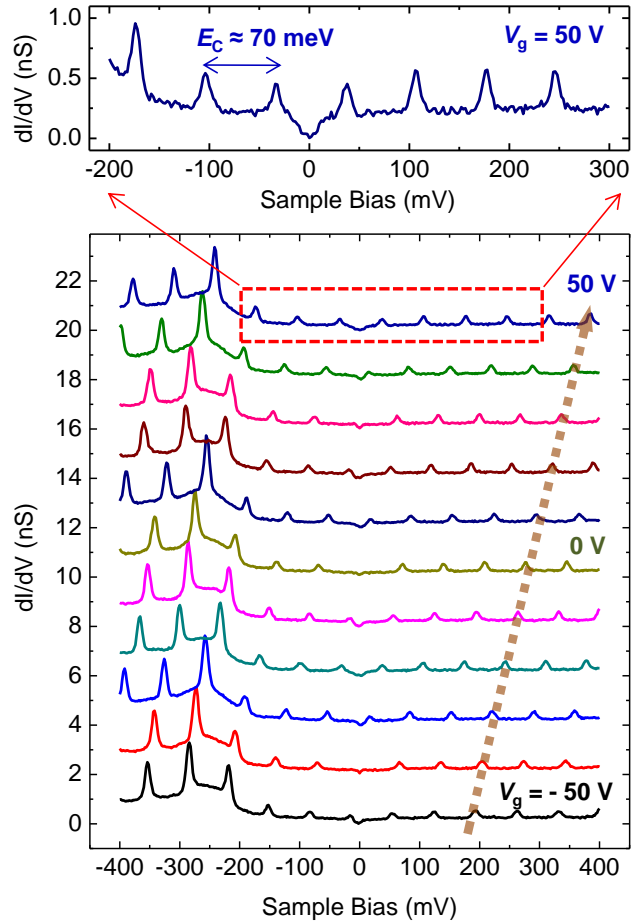

**Figure S6**: STS *dI/dV* spectra as a function of gate voltage with a QD attached to the STM tip. The STM tip picked up a QD, whose charging energy ($E_C$) is in the order of 70 meV, in the region between the Au electrodes and single layer region. The origin of the QD formation is unknown, but is most likely due to the pickup of a loosened Au particle. Coulomb oscillations exist in all of the measured gate and bias voltage ranges without revealing the electronic structure of graphene. The charging energy and peak positions show weak gate voltage dependence. This QD disappeared from the tip after further scanning. STS parameters: set-point current $I = 100$ pA, sample bias $V_b = 400$ mV, RMS modulation voltage 4 mV and magnetic field $B = 6$ T. Each *dI/dV* spectrum is offset for clarification.